\documentclass[sigplan,screen]{acmart}

\setcopyright{none}
\renewcommand\footnotetextcopyrightpermission[1]{} 

\begin{document}

\title{zkFi: Privacy-Preserving and Regulation Compliant Transactions using Zero Knowledge Proofs \\\vspace*{20pt} \Large  June 2023}

\author{Amit Chaudhary}
\affiliation{\country{}}
\email{amit.chaudhary.3@warwick.ac.uk}

\thispagestyle{plain}
\pagestyle{plain}

\begin{abstract}
We propose a middleware solution designed to facilitate seamless integration of privacy using zero-knowledge proofs within various multi-chain protocols, encompassing domains such as DeFi, gaming, social networks, DAOs, e-commerce, and the metaverse. Our design achieves two divergent goals. zkFi aims to preserve consumer privacy while achieving regulation compliance through zero-knowledge proofs. These ends are simultaneously achievable. zkFi protocol is designed to function as a plug-and-play solution, offering developers the flexibility to handle transactional assets while abstracting away the complexities associated with zero-knowledge proofs. Notably, specific expertise in zero-knowledge proofs (ZKP) is optional, attributed to zkFi's modular approach and software development kit (SDK) availability.
\end{abstract}

\begin{teaserfigure}
  \includegraphics[width=\textwidth]{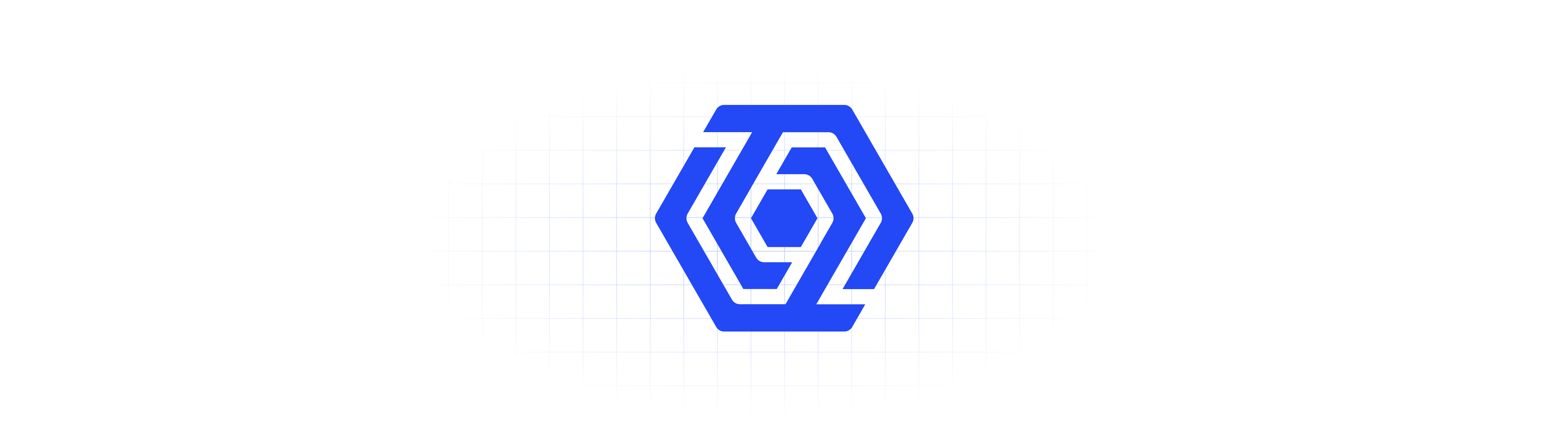}
\end{teaserfigure}

\maketitle

\section{Introduction}

Achieving privacy in blockchain applications presents unique challenges - often requiring trade-offs between user experience and privacy. The transparent nature of conventional blockchains reveals all of the transaction data, including addresses, assets involved, amount, smart-contract data, and timestamps, out to the public. It is analogous to using a regular bank account and revealing all private financial information, deterring the mass adoption of blockchain and digital asset technology.

As this space continues to evolve and more institutional and individual users engage in activities on these applications, privacy will become a paramount concern creating the biggest hurdle for achieving mainstream adoption. Individuals contemplating the adoption of blockchain-based payment systems may exhibit considerable hesitance if their salaries or other confidential financial details, such as payments for medical services and their online purchases, are accessible to the public. This demand for privacy will also be from social networking platforms, decentralized lending protocols, philanthropic platforms, e-commerce, gaming, and other protocols where users want to prioritize safeguarding the privacy of their information.

While there is a clear need for privacy solutions, regulatory scrutiny of privacy protocols necessitates action to develop practical and fair measures that deter bad actors from engaging in on-chain illicit activity. \textbf{Selective De-anonymization}, as mentioned in \cite{a16z}, lays out a method for allowing traceability. Particularly, an instantiation of \textbf{involuntary de-anonymization} as practically studied in \cite{compliance-sol} can prove to be a flagship regulation-compliant technique that can be used when a malicious actor refuses to comply with the law.

In this paper, we propose a privacy-preserving solution with solid regulatory compliance using zero-knowledge proofs and threshold cryptography having the following features:

\begin{itemize}
    \item A general purpose \textbf{multi-chain} privacy solution spanning across multiple EVM chains.
    \item Available with \textbf{simple, composable and flexible plug-and-play middleware} solution via an SDK.
    \item Secure with \textbf{built-in compliance} solution with concrete AML practices.
    \item Providing a seamless \textbf{user-experience} using account abstraction and wallet integrations \cite{mm-snaps, ledger-apps}.
\end{itemize}

\section{Limitations in current architecture}
At present most widely used programmable blockchains (e.g. EVM based chains such as Ethereum, Polygon, Optimism, Arbitrum) offer benefits such as permissionless nature, decentralization, and security, but these blockchains do not offer privacy. Alternative blockchain networks have been aiming to create solutions from scratch, to eliminate the pitfalls but fail to near the activity and value of mentioned public chains. This necessitates a solution to multiple problems on the public chain itself.

\paragraph{Lack of Privacy}
A regular on-chain transaction exposes private data and transactions to the public. The data such as sender/receiver address, asset type, and quantity, smart-contract data, timestamps, etc. are conveniently available in an organized manner to the general public through block explorers such as Etherscan \cite{etherscan}. This information can be used to track funds for targeted attacks, identify users, and extract sensitive information and patterns about their activity.
These pitfalls prevent the adoption of revolutionary blockchain applications by several serious users, especially users like institutional investors and businesses.

\paragraph{Weak Compliance}
Privacy problem implicitly poses another severe issue of compliance. How to have robust compliance in place and prevent bad actors while maintaining user privacy? Enabling privacy on decentralized platforms has been very well known to attract malicious actors abusing the platform. These include using it for illicit activities like laundering stolen funds from hacks or preparing for one. Most of the time, these actors have succeeded because of the lack of firm AML (Anti-Money Laundering) practices to deter bad actors. Weak compliance deters institutional investors or businesses from entering the blockchain space for legitimate usage.

\paragraph{Lack of Infrastructure}
Building private applications on the blockchain has been made possible by the advancements in Zero Knowledge (ZK). However, implementing ZK technology is complicated. Developers need specialized knowledge and resource investment in ZK development. This creates overhead and distraction from the divergence of resources from developing a core of their applications. 

\paragraph{Poor User-Experience}
These distractions and overheads due to the lack of middleware privacy solutions lead to a subpar developer and user experience. Even if the application develops its privacy layer, the UI/UX faces challenges for users. While UX is still an area to be improved in web3 generally - it is even worse in the ecosystem of privacy-preserving applications.

\subsection{Previous Solutions}

\textbf{ZCash} blockchain was among the first to tackle privacy by facilitating anonymous transactions. While innovations there have been impressive, they could not reach the intended adoption nor offered any programmability - restricting only to peer-to-peer transactions. Hence, missing out on the prominent application level use-case for, eg. DeFi. \textbf{Monero}, another private chain, more or less shares the same context.

\textbf{Tornado} protocol on Ethereum amassed a significant number of usage despite not-so-good UX. But it overlooked compliance and became go-to-place for money laundering \cite{trm-labs}. A portion of its volume has ties to large-scale hacks \cite{Chainalysis}, attracting serious implications from regulators. It, too, only offered peer-to-peer private transactions lacking any interoperability beyond that.
Aztec came up with a novel solution with their L2 roll-up approach that had DeFi compatibility to some extent. However, it required users to bridge their assets back and forth and had significant waiting times - making it not-so-practical for all kinds of DeFi interaction, e.g. in swaps because of slippage. 

Others also share the same problems, especially weak compliance guarantees and friction in user experiences.

\section{zkFi: A Middleware Solution}

\begin{figure*}
    \centering
    \includegraphics[scale=0.4]{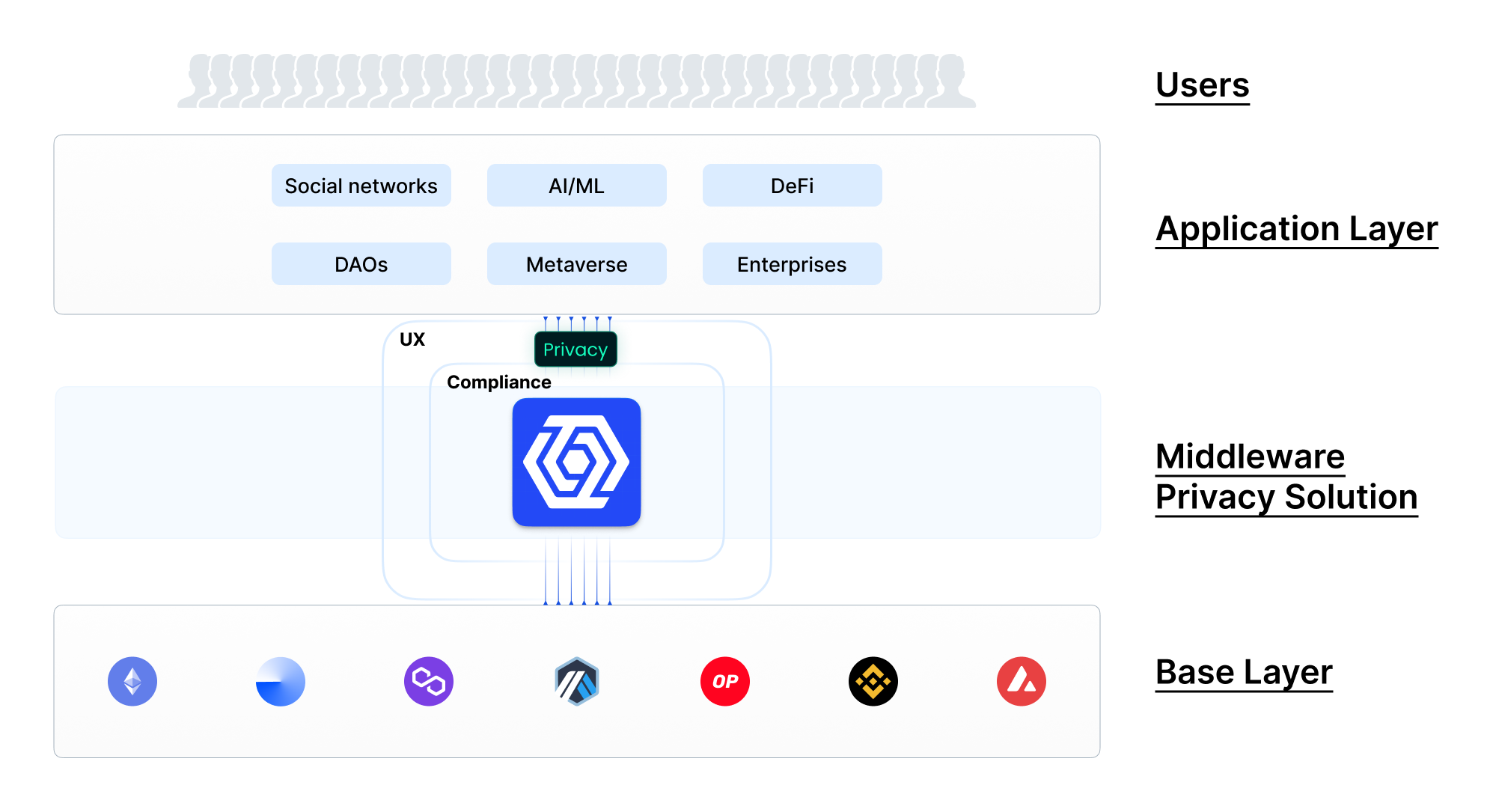}
    \caption{Middleware Solution}
    \label{fig:middleware}
\end{figure*}

To tackle problems, as discussed above, zkFi offers a packaged solution that acts as a privacy middleware with built-in compliance. Privacy and compliance-related complexity are abstracted away by providing the developers with an SDK that facilitates a plug-and-play solution.

\subsection{Privacy with Zero Knowledge Proofs (ZKPs)}
zkFi uses ZKPs to facilitate privacy at its core by achieving the following goals:
\begin{itemize}
    \item Performing private transactions concealing sender, recipient, and amount of funds being transferred.
    \item Ensuring transactions cannot be linked together, preventing tracking the flow of funds.
    \item Preventing double-spending by proving the transaction is valid without revealing information about the transaction itself.
    \item Selective de-anonymization by ensuring verifiable encryption of transaction data.
\end{itemize}

While there are multiple formulations of ZKP systems available and being researched, zkFi specifically utilizes a groth16 \cite{groth16} zkSNARK system which is currently most suitable for on-chain privacy applications.

\subsection{Stronger Compliance Guarantees}
Compliance has been the conundrum for privacy protocols so far. zkFi aims to have an industry-standard compliance framework such as: 

\begin{itemize}
\item \textbf{Selective De-Anonymization:}
For an industry-standard AML practice, a process for the de-anonymization of user-specific private transaction data needs to be followed. It could be a \textbf{voluntary de-anonymization} where the entity in question can share general or per-transaction viewing key to a regulatory authority. In other cases (for malicious actors), \textbf{involuntary de-anonymization} may be enforced in response to a regulatory emergency or court order. The latter is an accountable multi-party process to prevent abuse of power. It is thoroughly studied in \cite{compliance-sol} as SeDe (short for Selective De-anonymization) framework for compliance and is a flagship among all compliance tools available.

\item \textbf{Deposit Limits:}
We put a fiat limit on the asset being transacted and/or the volume flowing through the protocol in a time period. There can be a provision to relax this limit for specific entities like a compliant businesses.


\item \textbf{Risk Management and Screening}
The protocol sets up compliance and risk management integrations to identify and prevent any kind of illegal financial activity. This could be achieved by services like {TRM Labs}\footnote{https://www.trmlabs.com/} products and Chainalysis\footnote{https://go.chainalysis.com/chainalysis-oracle-docs.html} oracles perform screening and identifying inflow of illegal funds into the system.

\end{itemize}

\subsection{Pluggable Privacy With SDK} \label{sdk}

By offering an SDK, a full set of compliant privacy features is instantly available to protocols and their developers. SDK facilitates a \textbf{simple composable plug-and-play solution} that abstracts away every bit of ZKPs or compliance-specific complexities. This renders immense benefits to protocols such as:
\begin{itemize}

    \item New protocols can focus on developing their core features without investing time and resources into ZK development.
    
    \item Existing protocols do not have to modify their smart contracts for compatibility. It's a simple plug-and-play through SDK.
    
    \item Compliance issues often come as part of a privacy conundrum. But with zkFi, protocols will not have the burden of juggling privacy-related compliance practices.
\end{itemize}

In addition to the advantages mentioned above, the \textbf{flexibility of interaction} with the integrating protocol remains at the hands of a developer. Given private user assets and/or data as input, developers are free to write any custom logic to apply to the inputs.

See section \ref{proxy} for implementation details.

\subsection{Account Abstraction and UX}
The advent of EIP-4337 \cite{eip-4337} Account Abstraction proposal allowed for significant improvements in user experience across protocols. One such improvement or feature it brings is the ability for smart contracts to pay for gas fees of transactions. This allows for gas-sponsored transactions or payment of gas fees in ERC-20 tokens.

In privacy protocols, there exists a common problem referred to as \textit{gas dilemma} or \textit{fee payment dilemma}. A gas dilemma exists because if users need to pay the gas fees from their wallet to execute their transactions then this gas payment discloses the user's public profile since their address as the sender of the transaction is now visible publicly.

zkFi uses account abstraction features from EIP-4337 so that the gas fee can be paid in ERC-20 tokens making transactions on zkFi a smooth experience. The flexibility of gas payment is left to the integrating protocol - allowing it to sponsor transaction fees or charge from their users in any desired way. In the case of peer-to-peer transactions, a user simply pays gas in transacting assets. This is facilitated through a custom EIP-4337 \textbf{Paymaster} contract that pays the gas on the user's behalf in exchange for a small fee in any supported asset.

\subsection{Wallet Integrations} \label{wallet-integration}
By implementing a simple interface, provided within SDK, for the shielded account operations (e.g. signing private transactions), crypto wallet applications can support private transactions for their users. In that case, the shielded account keys share the same security space as the wallet's private keys. Developers simply send defined requests to the shielded account for, say, authorizing transactions by signing it. This allows developers to directly use the shielded account in their application UIs without concern about securely handling sensitive keys.

Some examples of such wallet integrations are:

\begin{itemize}
\item \textbf{MetaMask}: MetaMask is a browser extension as a crypto wallet. Through its \textbf{Snaps API} \cite{mm-snaps} it allows developers to extend its functionality by publishing custom packages of logic called "Snap". One such Snap, the "zkFi Snap" acts as an interface to the shielded accounts within MetaMask.

\item \textbf{Ledger}: Ledger is a hardware wallet that securely stores private keys for multiple cryptocurrencies. It allows \textbf{Embedded Apps} \cite{ledger-apps} to be installed on the user's device. A zkFi app is one such app tailored to securely store shielded account keys and authorize private transactions from the device.
\end{itemize}

\section{Use Cases of zkFi}

\paragraph{Pluggable Privacy for DeFi protocols}
As mentioned, previously in section \ref{sdk}, the design of the infrastructure allows any existing DeFi protocol for seamless integration to enable anonymous transactions.

For instance, consider Aave\footnote{https://aave.com/} which is a lending/borrowing liquidity protocol. Aave users normally supply assets to one of Aave markets and get interest-bearing tokens a.k.a \textit{aToken}\footnote{https://docs.aave.com/developers/tokens/atoken} in return. For Aave to let its users supply assets anonymously, it'd just require a simple stateless proxy contract, let's call it \textit{AaveProxy}. The job of \textit{AaveProxy} is just to take an asset and return the corresponding \textit{aToken} asset by talking to APIs already written by Aave in their core contracts. The proxy is very loosely coupled, no ZK circuit programming is required at all nor any changes in Aave's existing contracts.

The same kind of seamless integration is possible with other protocols for - earning interest anonymously on Compound, doing anonymous swaps on Uniswap, staking anonymously on Lido, and many more.

\paragraph{Private Payments via Stealth Address}
While normal peer-to-peer private transactions are supported, one may opt to receive assets in the form of payment to their stealth address - which preserves anonymity one can share a random-looking address each time.

So, for instance, \textbf{payment links} can be generated by a receiver and can be shared with a sender such that the shared link has no traceability to any previous payment. This is similar to sharing payment links in Stripe\footnote{https://stripe.com/in/payments/payment-links} but with anonymity.

\paragraph{Shielded Account for protocol UIs}
Via its integration directly in existing crypto wallets, starting with MetaMask Snaps integration, any protocol may request spending of user's private assets instead of public assets from a normal wallet. This frees protocols to define their own custom UIs on their own domains and request interaction with Shielded Wallet however it wants.

This adds up with a better user experience thanks to built-in account abstraction features like gasless transactions.
\\
\\
zkFi strives to offer a general solution, so it is future compatible with other use cases that may arise as the need for privacy is realized more and more.

\section{Architecture}
The diagram in figure \ref{fig:architecture} shows an architectural overview of the system with involved actors and their interaction with each other:

\begin{figure*}
    \centering
    \includegraphics[scale=0.20]{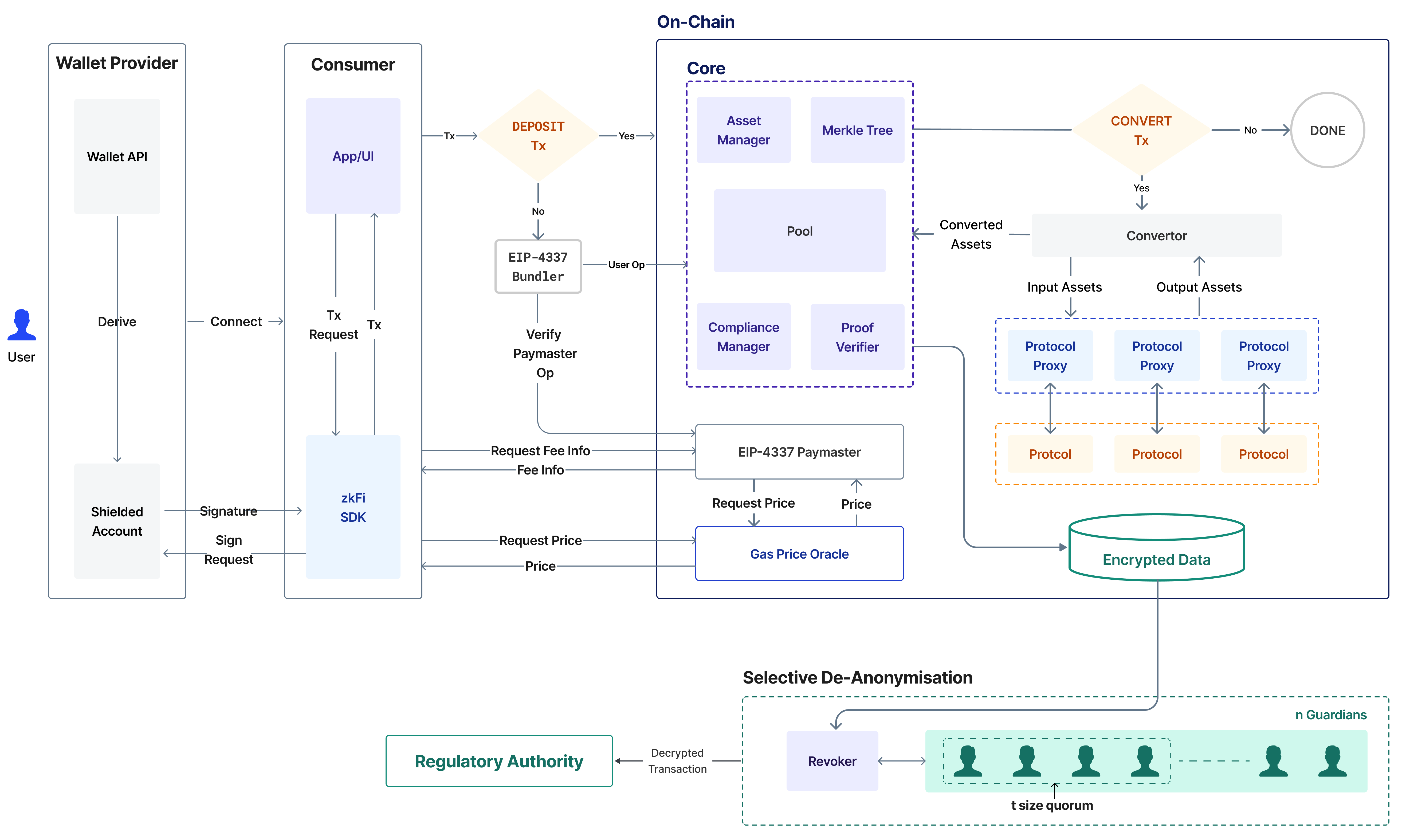}
    \caption{Architecture Diagram}
    \label{fig:architecture}
\end{figure*}

\paragraph{Wallet Provider} The host for shielded account. This includes crypto wallets (e.g. MetaMask and Ledger) which provide functionalities to deterministically derive keys of shielded account from ethereum account of user.

\paragraph{Consumer} The consumer of the SDK that communicates with the Shielded Account via a connection to the wallet provider. It can invoke private transaction approvals by requesting transaction signature. The SDK consumers can be an application user interface or command-line applications which can construct complex transactions by passing only simple transaction requests to SDK. The SDK parses the request and handles creation, signing and (ZK) proving of the transaction. Ultimately, SDK outputs a suitable transaction payload to be sent on-chain.

\paragraph{EIP-4337 Bundler} A bundler sends the transactions to the network through a EIP-4337 bundler node rather than directly from a wallet. This has two main benefits - the user avoids exposing their wallet address publicly and it can pay for the gas with shielded/private assets itself.

\paragraph{Gas Price Oracle} A decentralized oracle to consulted by SDK and paymaster to calculate the equivalent amount of gas price in ERC-20 tokens at the time of transaction. A decentralized exchange like Uniswap \cite{uniswap-oracle} could be a suitable oracle.

\paragraph{EIP-4337 Paymaster} A custom EIP-4337 compatible paymaster that pays for the user operations in exchange for fees cut in the asset being transacted. The paymaster validates the operation after consulting Gas Price Oracle and making sure enough fee will be paid.

\paragraph{Core} The core of zkFi protocol encapsulates a \textbf{multi-transactional multi-asset} pool, an asset manager, a merkle tree of notes and an on-chain ZK proof verifier. See section \ref{core}.

\paragraph{Convertor} A smart contract that mediates the \textbf{convert} operation on behalf of core. It involves calling target protocol (e.g. DeFi) via its proxy for performing an operation (e.g. swap, stake, lend) and returning any resulting assets as a result of operation, back to core.

\paragraph{Protocol Proxy} Proxies are simple smart contracts that implement/extend a simple interface/base provided by SDK. Doing this allows it to plug nicely into the Core and be able to receive assets from it to perform any DeFi operation. See section \ref{proxy}.

\paragraph{Guardians} Guardians are multi-party entities that exist to perform the \textbf{involuntary selective de-anonymization} process upon a verifiable request from a revoker entity as per the \cite{compliance-sol}. See section \ref{deanon}.

\section{Building Blocks}

\subsection{Shielded Account}
A user holding a shielded account can sign valid transactions and decrypt balances and transaction data. Unlike normal Ethereum accounts which are controlled by a single private key, a shielded account contains two types of such keys.

We assume the existence of a cryptographically secure random number generator (RNG) function used to generate private keys:

\begin{equation} \label{eq:rng}
    \mathtt{RNG} \colon \phi \to \mathbb{B}^{256} ; () \mapsto \xi
\end{equation}

We cryptographically derive the keys of the shielded account using the $\mathtt{poseidon}$ hash function (from \ref{eq:poseidon}) and group operations on the BabyJubJub curve (see Sec. \ref{ec}).

\subsubsection{Sign Key}
The sign private key can sign valid transactions authorizing the spending of user funds in shielded accounts. A new private key is generated by first sampling an entropy element, $\xi$ by invoking $\mathtt{RNG}$ and then hashing it with a fixed salt value $\Delta_{sign}$.

\begin{equation}
s = \mathtt{poseidon}(\xi, \Delta_{sign})
\end{equation}

The corresponding public key $S$ is simply a point on the curve as:
\begin{equation}
    S = s \cdot G
\end{equation} 

\subsubsection{View Key}

The view private key decrypts the user balances and transaction history. This private key is generated by hashing the entropy $\xi$ with salt $\Delta_{view}$:
\begin{equation}
    p = \mathtt{poseidon}(\xi, \Delta_{view})
\end{equation}

and public key is $P = p \cdot G$.

During the construction of a transaction, $p$ is used to symmetrically encrypt transaction data (e.g. amounts, owner) using \textbf{ChaCha20-Poly1305} encryption scheme \cite{chacha20}. The resulting ciphertext is decrypted and later utilized for future transactions.

Having a distinct key for read-only access renders multiple advantages including revealing $p$ for compliance purposes without giving up spending authority and allowing protocol websites for read-only access of data to display on their custom UIs.

This also allows to have a transaction-specific view key giving the ability to reveal only selected transactions or protocol-specific viewing keys so that protocol websites only get to read data relevant to it rather than the entire transaction history.

\subsubsection{Shielded Address}

The shielded address, $A$ of a shielded account is the public key of both key pairs:
\begin{equation}
    A \equiv (S, P)
\end{equation}

A user may choose to post $A$ on a public address registry for others to look it up using public wallet addresses or ENS names and conveniently send funds to address $A$. Although $A$ is not directly used as is while associating funds to it in a zkFi transaction, it allows the derivation of a random-looking "stealth address", $x$ from $A$. $x$ is then used instead for the purpose.

\bigbreak

For the sake of better user experience for crypto wallet users, the $\mathtt{RNG}$ at equation \ref{eq:rng} should be facilitated by the wallet application, and the entropy $\xi$ be seeded by the private key or seed phrase. This is possible with the wallet integrations (see Sec \ref{wallet-integration}). In that case, the keys of the shielded account can be derived from the wallet account itself. This frees the user from handling additional keys. 

\subsection{Stealth Address}
A stealth address is a one-time address that is randomly generated to receive funds to. This address is not linked to the receiver's permanent address. Hence making it difficult to link or track a particular user's transactions and protect their identity.

There are multiple different schemes available for deriving a stealth address from a user account. These schemes allow the receiver to detect funded stealth addresses meant for them and calculate the associated private key from auxiliary data, normally sent along with the transaction. The calculated private key can then be used to sign transactions authorizing the spending of funds at that stealth address. One example of such a scheme is defined in EIP-5564 \cite{eip-5564}.

In zkFi, we primarily use ZK proofs (but also signatures) for constructing valid transactions. This allows for a much simpler and more efficient way to generate and use stealth addresses in the protocol. A zkFi stealth address, $x$ is derived from sign public key $S$ as:
\begin{equation} \label{eq:stealth-address}
    x = \mathtt{poseidon}((S, \delta))
\end{equation}

where $\delta$ is a random element also referred to as a blinding factor.

To relay any auxiliary data to be included in the transaction request, the sender generates an ephemeral public key, $Q = r \cdot G$ where $r \in \mathbb{B}^{248}$ is another random element.

A sender then calculates a shared key $K$ from the view public key $P$ of the receiver as:
\begin{equation} \label{eq:shared-key}
    K = r \cdot P
\end{equation}

$K$ is used to encrypt any desired sensitive transaction data, e.g. $\delta$, to get a ciphertext, $C$.

A \textbf{view tag}, $t$ for this stealth address or transaction is the most significant byte of the hash of $K$:
\begin{equation} \label{eq:view-tag}
    t = \mathtt{keccak256}(K)[0:1]
\end{equation}

$t$, $x$, $Q$ and $C$ are concatenated to form auxiliary data $(t \Vert x \Vert Q \Vert C)$ to be broadcasted along with the transaction.

A user parses the auxiliary data of a transaction and tries to calculate a shared key $K^{\prime}$ using its view private key $p^{\prime}$ and $Q$ as:
\begin{equation}
    K^{\prime} = p^{\prime} \cdot Q
\end{equation}

And calculates a view tag $v^{\prime}$ from $K^{\prime}$ using equation \ref{eq:view-tag}. If $v^{\prime} \neq v$, the user is not the receiver. It stops.

Otherwise, the receiver continues and uses $K^{\prime}$ to perform decryption of $C$ to retrieve the plaintext data, e.g. $\delta^{\prime}$, that was encrypted before. The receiver uses $\delta^{\prime}$ to calculate the stealth address $x^{\prime}$ (using equation \ref{eq:stealth-address}). The user performs an additional check that $x^{\prime} = x$ (hence, $\delta^{\prime} = \delta$) to be sure that it is indeed the receiver just in case the view tag match was a false positive.

To spend the asset the receiver is now able to prove a statement $\mathcal{S}_{x}$ defined as:
\begin{equation}
    \mathcal{S}_{x} \equiv \text{Knowledge of } (s, \delta) \colon x = \mathtt{poseidon}(S, \delta)
\end{equation}

The proof of $\mathcal{S}_{x}$ is included in the ZK proof $\boldsymbol{\pi}$ of the transaction to prove the ownership of $x$.

\subsection{Core} \label{core}

Core smart contracts of the zkFi protocol. The core includes a \textbf{multi-transactional multi-asset pool}, meaning the pool supports multiple assets and can transact multiple assets in a single transaction. The on-chain ZK Verifier verifies the proof submitted during the transaction.

\subsubsection{Setup}
Let $\mathcal{T}$ be the Merkle tree whose leaves are calculated using the poseidon hash function (equation \ref{eq:poseidon}). The tree leaf nodes are subsequently filled with note $\mathbf{commitment}$s in an append-only fashion.

A note is a tuple $N$ of multiple elements:
\begin{equation}
N \equiv (e, x, v, \delta)
\end{equation}

where, $e \in \mathbb{B}^{24}$ is the asset identifier and $x \in \mathbb{B}^{248}$ is the stealth address of the owner, $v \in \mathbb{B}^{248}$ is associated value of the note and $\delta$ is the blinding factor from which $x$ was generated.

The $\mathbf{commitment}$, $c$ of a note is:
\begin{equation}
c = \mathtt{poseidon}(e, x, v)
\end{equation}

Let $\sigma$ be the signature generated with sign key $s$ that authenticates the ownership of $N$ as:
\begin{equation}
    \sigma = \mathtt{Schnorr\_Sign}(c, s)
\end{equation}

where $\mathtt{Schnorr\_Sign}$ is as defined in \ref{schnorr-sign}.

Let $\eta$ be the $\textbf{nullifier}$ hash of note $N$ defined as:
\begin{equation}
    \eta = \mathtt{poseidon}(l, c, \delta)
\end{equation}

where $l$ is index of note commitment $c$ as leaf node in $\mathcal{T}$.

Let's define a set of \textbf{public inputs}, $\rho$ to the prover as:
$$
\rho \equiv (R, \mathbf{V}, \mathbf{E}, \boldsymbol{\eta}^{in}, \mathbf{c}^{out})
$$

where, 
\begin{itemize}
    \item $R$ is root of $\mathcal{T}$.
    \item $\mathbf{V}$ is a list of $m$ public values for each output note. A value $V$ in list $\mathbf{V}$ is considered negative for any value leaving the pool, positive otherwise.
    \item $\mathbf{E}^{in}$ is a list of public asset identifiers corresponding to public values.
    \item $\boldsymbol{\eta}^{in} = (\eta_{1}, \eta_{2},..., \eta_{n})$ is list of nullifier hashes corresponding to $n$ input notes $\mathbf{N}^{in} = (N^{in}_1, N^{in}_2, ...N^{in}_n)$.
    \item $\mathbf{c}^{out}$ is list of commitments of $m$ output notes $\mathbf{N}^{out} = (N^{out}_1, N^{out}_2, ..., N^{out}_m)$.
\end{itemize}

Similarly, we define a tuple of \textbf{private inputs} as:
\begin{equation}
\omega \equiv (\mathbf{e}^{in}, \mathbf{e}^{out}, \mathbf{v}^{in}, \mathbf{v}^{out}, \boldsymbol{\delta}^{in}, \mathbf{S}^{in}, \boldsymbol{\sigma}^{in}, \mathbf{x}^{out}, \mathbf{l}^{in}, \mathbf{o}^{in})
\end{equation}

where

\begin{itemize}
    \item $\mathbf{e}^{in}$ and $\mathbf{e}^{out}$ are the lists of asset identifiers of input notes and output notes.
    
    \item $\mathbf{v}^{in} = (v^{in}_1, v^{in}_2, ..., v^{in}_n)$ and $\mathbf{v}^{out} = (v^{out}_1, v^{out}_2, ..., v^{out}_m)$ are the values of input notes, $\mathbf{N}^{in}$ and output notes in $\mathbf{N}^{out}$.
    
    \item $\boldsymbol{\delta}^{in}$ is a list of blinding factors of stealth addresses attached to input notes, $\mathbf{N}^{in}$.
    
    \item $\textbf{S}^{in} = (S^{in}_1, S^{in}_2, ...,S^{in}_n)$ are public keys associated with $n$ input notes $\textbf{N}^{in}$.
    
    \item $\boldsymbol{\sigma}^{in}$ are the signatures produced by signing commitments of $\textbf{N}^{in}$ with private keys $\textbf{s}^{in}$ corresponding to public keys $\textbf{S}^{in}$.
    
    \item $\mathbf{x}^{out}$ are stealth addresses associated with output notes $\mathbf{N}^{out}$.
    
    \item $\mathbf{l}^{in}$ are leaf indices of input note commitments, $\mathbf{c}^{in}$.
    
    \item $\mathbf{o}^{in}$ are openings in $\mathcal{T}$ at indices $\mathbf{l}^{in}$ respectively.
\end{itemize}

Then, given $\rho$ and $\omega$, let $\mathcal{S}$ be the statement of knowledge defined as:
\begin{equation}
\begin{aligned}
\mathcal{S} \equiv \ & \text{Knowledge of}  \ \omega \colon \\ 
  & \ \ \ \forall i \in [1, n], \ x^{in}_i = \mathtt{poseidon}(S^{in}_i, \delta^{in}_i) \\
  & \wedge \forall i \in [1, n], \ c^{in}_i = \mathtt{poseidon}(e^{in}_i, x^{in}_i, v^{in}_i) \\
  & \wedge \forall i \in [1, n], \ \eta^{in}_i = \mathtt{poseidon}(l^{in}_i, c^{in}_i, \delta^{in}_i) \\
  & \wedge \forall i \in [1, n], \ \mathtt{Schnorr\_Verify}(c^{in}_i, S^{in}_i, \sigma^{in}_i) = 1 \\
  & \wedge \forall i \in [1, n], \ o^{in}_i \ \text{is the opening at} \ l^{in}_i \ \text{in} \ \mathcal{T} \\
  & \wedge \forall i \in [1, m], \ {c}^{out}_i = \mathtt{poseidon}(e^{out}_i, x^{out}_i, v^{out}_i) \\
  & \wedge \forall e^{in}_i \in \mathbf{e}^{in}, \sum \mathtt{selv}(\mathbf{v}^{in}, e^{in}_i) + \sum \mathtt{selv}(\mathbf{V}, e^{in}_i) \\ & \hspace{80pt} = \sum \mathtt{selv}(\mathbf{v}^{out}, e^{in}_i) \\
  & \wedge \forall \ e^{out}_i \in \mathtt{sele}(\mathbf{E}, \mathbf{e}^{out}), \sum \mathtt{selv}(\mathbf{v}^{in}, e^{out}_i) \\ & \hspace{30pt} + \sum \mathtt{selv}(\mathbf{V}, e^{out}_i) = \sum \mathtt{selv}(\mathbf{v}^{out}, e^{out}_i)    
\end{aligned}
\end{equation}

where $\mathtt{selv}$ is a function to filter select values from a list of values by a given asset identifier:

\begin{equation}
\mathtt{selv}(\mathbf{v}, e) = \{ v \colon v \in \mathbf{v} \ \land v \text{ is the value of asset } e \}
\end{equation}

and $\mathtt{sele}$ is a function to select elements from the first list parameter if non-zero, otherwise selecting from the second list parameter:
\begin{equation}
\mathtt{sele}(\mathbf{E}, \mathbf{e}) = \{ e \colon e = E_i \text{ if } E_i \neq 0, \text{ otherwise } e = e_i \}
\end{equation}

$\mathtt{Schnorr\_Verify}$ is defined in \ref{schnorr-verify}.

Let $d_p$ and $d_v$ be the proving and verifying keys created using a trusted setup ceremony. Let's define the SNARK proof generator function as:
\begin{equation}
\mathtt{Prove} : \mathbb{B}^* \to \mathbb{B}^{2048} ; (d_p, \rho, \omega) \mapsto \boldsymbol{\pi}
\end{equation}
where $\boldsymbol{\pi}$ is called the \textbf{proof}.

And proof verifier as:
\begin{equation}
\mathtt{Verify} : \mathbb{B}^* \to \{0, 1\} ; (d_v, \boldsymbol{\pi}, \rho) \mapsto y
\end{equation}
where $y$ is a single bit for representing boolean $true$ or $false$ as a result of proof verification.

\subsubsection{Creating Transaction}
A transaction request is constructed by following the steps below:

\begin{enumerate}
    \item Identify the different assets $\mathbf{e} = (e_1, e_2, ...e_n)$ needed for transaction.
    \item Scan the network and fetch encrypted notes data. Decrypt it using the viewing key and filter for notes, $N_i \in \mathbf{N}$ such that its asset id, $e \in \mathbf{e}$.
    \item Choose a number of notes $\mathbf{N}^{in} \subset \mathbf{N}$, with commitment indices $\mathbf{l}_{in}$ in $\mathcal{T}$, to spend.
    \item Compute nullifier hashes, $\mathbf{\eta}^{in}$ of notes $\mathbf{N}^{in}$.
    \item Select a recent root, $R$ of $\mathcal{T}$ and calculate tree openings $\mathbf{o}^{in}$ at indices $\mathbf{l}^{in}$.
    \item Derive stealth addresses $\mathbf{x}^{out}$ of recipients and create output notes $\mathbf{N}^{out}$ associated with them. Also, calculate output commitments, $\mathbf{c}^{out}$.
    \item Set $\mathbf{V}$ and $\mathbf{E}$ to control any public asset value, if any, going into or coming out of contracts.
\end{enumerate}

\subsubsection{Signing Transaction}
To authorize the transaction, signatures $\mathbf{\sigma^{in}}$ from the shielded account are required. Each signature $\sigma^{in}_i \in \mathbf{\sigma^{in}}$ is obtained by signing $c^{in}_i$ of notes being spent ($\mathbf{N}^{in}$):

\begin{equation}
    \sigma^{in}_{i} = \mathtt{Schnorr\_Sign}(c^{in}_i, s)
\end{equation}

\subsubsection{Proving Transaction}
Together with signatures and parameters identified or calculated during creation, the public and private inputs i.e. $\rho$ and $\omega$ are put together. Then proof $\boldsymbol{\pi}$ of the transaction generated:

\begin{equation}
    \boldsymbol{\pi} = \mathtt{Prove}(d_p, \rho, \omega)
\end{equation}

\subsubsection{Sending Transaction}
Except for the deposit transactions type (public value going into pool), the transaction is sent as an EIP-4337 user operation \cite{eip-4337} by submitting it to a bundler node.

For deposits, it is sent directly from a wallet since it involves transferring value from the wallet to the pool smart contract.

If $\boldsymbol{\pi}$ was calculated correctly and the statement was true, the on-chain $\mathtt{Verify}$ function outputs $1$, and flow proceeds with necessary funds. Otherwise, $0$ representing bad or tampered $\boldsymbol{\pi}$.
\bigbreak

\subsection{Protocol Proxy} \label{proxy}
A \textbf{Protocol Proxy} is a state-less contract that lives on-chain to act as a proxy for a target (e.g. DeFi) protocol. This contract must implement a standard interface that exposes a "\textbf{convert}" functionality or operation. This reflects the fact that, in general, any DeFi operation e.g. swapping, lending, staking, can be thought of as a process of converting from one asset called \textit{input asset} to another asset called \textit{output asset}. Here are a few examples:
\begin{itemize}
    \item When swapping ETH for USDC on Uniswap, ETH (input asset) is converted to USDC (output asset).
    \item When supplying DAI on the Aave market to earn interest, DAI (input asset) is essentially being converted to aDAI (output asset) the interest-bearing tokens given back in return.
    \item Staking ETH (input asset) on Lido results in conversion to stETH (output asset) holding which represents your stake plus rewards.
\end{itemize}

This allows \textbf{Core} the core to interact with protocols via a \textbf{Convertor} contract. In its \textbf{convert} operation, this proxy is supposed to perform necessary operations by calling the target it is the proxy for, eventually returning any output to the Core.

It is a specific kind of transaction where $V$ values are set to non-zero so that these values, $\mathbf{v}$ are sent as input asset (with ids $\mathbf{e}$) if required. A fee asset of id $e_f$ and value $v_f$ is also given specifying gas fee requirements. Fee must be paid by returning it as one of the output assets. The target protocol may chose to pay gas however it wants. Lastly, the input also includes an arbitrary payload, $d$ necessary for operation by the proxy. After processing the input, $(\mathbf{e}, \mathbf{v}, e_f, v_f, d)$ with its specific target protocol, the proxy is expected to return output asset  $(\mathbf{e}^{\prime}, \mathbf{v}^{\prime})$.
\begin{equation}
\begin{split}
\mathtt{Convert} \colon & ({\mathbb{B}^{24}}^n, {\mathbb{B}^{256}}^n, \mathbb{B}^{24}, \mathbb{B}^{256}, \mathbb{B}^{*}) \to ({\mathbb{B}^{24}}^m, {\mathbb{B}^{256}}^m) \\
& ; (\mathbf{e}, \mathbf{v}, e_f, v_f, d) \mapsto (\mathbf{e}^{\prime}, \mathbf{v}^{\prime})
\end{split}
\end{equation}

Later, a note commitment is calculated with $(e^{\prime}, v^{\prime})$ for each of $m$ outputs and a given stealth address as a transaction parameter and inserted into $\mathcal{T}$.

The flexibility comes from the fact that the implementation of $\mathtt{Convert}$ is left to the developer trying to integrate the target protocol.

\subsection{Involuntary Selective De-anonymization} \label{deanon}
zkFi adopts a framework for involuntary de-anonymization of illicit transaction subgraphs using a threshold multi-party procedure while keeping such parties accountable.

The process involves an independent set of parties called guardians. Each of the $n$ guardians holds a share of the secret key which decrypts encrypted transaction data such that guardians must reach a quorum of minimum size $t$ ($t \leq n$) for the de-anonymization of any transaction to occur. However, even after de-anonymization is granted by guardians, they still do not learn any information about the user because another decryption still needs to be performed by an individual party called the revoker.

Accountability comes from the fact that the revoker is bound to send a publicly verifiable request of de-anonymization to the guardians who must agree first for the revoker to extract any information at all. A more detailed study can be found in \cite{compliance-sol}.

\section{Cryptographic Primitives}

\subsection{Hash Function}
A hash function is a mathematical function, $h$ that takes arbitrary length input, $m \in \mathbb{B}^*$, and outputs fixed-length output $x \in \mathbb{B}^n$ often called hash, hash value or digest.
\begin{equation}
    h \colon \mathbb{B}^* \to \mathbb{B}^n ; m \mapsto x
\end{equation}

A cryptographically secure hash function has the following properties:
\begin{itemize}
    \item \textbf{Pre-image resistance} Given a hash value, it must be difficult to find the input that produced it.
    \item \textbf{Second pre-image resistance} Given an input and its hash value it must be difficult to find another input that produces the same hash.
    \item \textbf{Collision resistance} It must be difficult to find two inputs that map to the same hash value.
\end{itemize}

Specifically for its ZK-related operations, zkFi extensively uses a hash function called \textbf{Poseidon} \cite{poseidon} denoted as $\mathtt{poseidon}$:
\begin{equation} \label{eq:poseidon}
    \mathtt{poseidon} \colon \mathbb{B}^* \to \mathbb{B}^{254}
\end{equation}

A common in the Ethereum ecosystem is the Keccak256 \cite{keccak256}. It is used within zkFi where it does not require proving via ZK proof. It is defined as:
\begin{equation} \label{eq:keccak256}
    \mathtt{keccak256} \colon \mathbb{B}^* \to \mathbb{B}^{256}
\end{equation}

\subsection{Elliptic Curve} \label{ec}
Elliptic curves are mathematical structures that are defined over some finite field that have interesting properties that make them very useful in cryptography. The security of ECC relies on the computational infeasibility of solving the elliptic curve \textbf{discrete logarithm problem (DLP)} \cite{dlp}. 

Let $E$ be an elliptic curve defined over a finite field $\mathbb{F}_q$ where $q$ is a large prime number, and let $P$, $Q$ and $R$ be points on $E$. The elliptic curve group operation on $E$ is typically denoted as $P + Q$, where  $P+Q+R$ satisfies the group law:

\begin{equation}
    P + Q + R = \mathcal{O}
\end{equation}

where $\mathcal{O}$ represents the point at infinity.

The elliptic curve DLP is defined as finding an integer $k$ such that $S = k \cdot P$, where $k \cdot P = P + P + .... + P$ ($k$ times) is referred to as scaler multiplication of point $P$.

A cyclic subgroup is a subset of an elliptic curve group that is generated by a single point called \textbf{generator}, denoted by $G$. Elements of the subgroup are obtained when $G$ is scalar multiplied by integers i.e. $k \cdot G$ where $k \in \mathbb{Z}$.

zkFi protocol uses a specific elliptic curve called \textbf{Baby JubJub} as defined in \cite{eip-2494}. It is particularly well-suited for cryptography operations in zero-knowledge (ZK) applications.

\subsection{Digital Signature}
A digital signature $\sigma$ is a verifiable piece of data produced by signing a message $m$ with a private key $s$ through some chosen signature scheme or function.
\begin{equation}
    \mathtt{Sign} \colon (\mathbb{B}^*, \mathbb{B}^k) \to \mathbb{B}^n ; (m, s) \mapsto \sigma
\end{equation}

The signature scheme can later verify that the signature $\sigma$ was produced by an entity who knows the $m$ as well as private key $s$:
\begin{equation}
    \mathtt{Verify} \colon (\mathbb{B}^n, \mathbb{B}^j) \to \{0, 1\} ; (\sigma, P) \mapsto y
\end{equation}
where $P$ is public key corresponding to $s$ and $y$ is a single bit representing result of verification - $true$ or $false$.

A cryptographically secure signature scheme must have the following properties:
\begin{itemize}
    \item \textbf{Authenticity} A valid signature implies that the message was deliberately signed by the signer only.
    \item \textbf{Unforgeability} Only the signer can produce a valid signature for the associated message.
    \item \textbf{Non-reusability} The signature of a message cannot be used on another message.
    \item \textbf{Non-repudiation} The signer of the message cannot deny having signed the message with a valid signature.
    \item \textbf{Integrity} Ensures that the contents of the message are not altered.
\end{itemize}

zkFi utilizes \textbf{Schnorr} \cite{schnorr} signature scheme for signing private transactions and verifying those signatures. The signing function is defined as:
\begin{equation} \label{schnorr-sign}
    \mathtt{Schnorr\_Sign} \colon (\mathbb{B}^{256}, \mathbb{B}^{512}) \to \mathbb{B}^{768} ; (m, s) \mapsto \sigma
\end{equation}

and verification as:
\begin{equation} \label{schnorr-verify}
    \mathtt{Schnorr\_Verify} \colon (\mathbb{B}^{256}, \mathbb{B}^{512}, \mathbb{B}^{512}) \to \{0, 1\} ; (m, S, \sigma) \mapsto b 
\end{equation}

where $b$ is a single bit representing result of verification $true$ or $false$.

\subsection{Zero Knowledge Proofs}
Zero-knowledge proof (ZKP) is a cryptographic technique that allows a party, a prover to prove to another party, a verifier that it knows a secret without actually revealing a secret by following a set of complex mathematical operations. 

Although the origins of ZKPs can be traced back to the early 1980s, when a group of researchers, including Shafi Goldwasser, Silvio Micali, and Charles Rackoff, published a paper \cite{zkp} that introduced the primitive concept to the world, it lacked practicality at the time. However, later developments addressed the problems leading to usable implementations in various systems, especially with the advent of blockchains.

In order to be ZKP, it must satisfy three properties:
\begin{itemize}
    \item \textbf{Completeness}: If the statement is true, both the prover and verifier are following the protocol, then the verifier will accept the proof.
    \item \textbf{Soundness}: If the statement is false, no prover can convince the verifier that it is true with any significant probability.
    \item \textbf{Zero Knowledge}: If a statement is true, the verifier learns nothing other than the fact that the statement is true.
\end{itemize}

One particularly efficient type of ZKP is \textbf{zkSNARK}. A \textbf{SNARK} (Succinct Non-Interactive Argument of Knowledge) defines a practical proof system where the proof is \textbf{succinct} that can be verified in a short time and is of small size. The system is \textbf{non-interactive} so that the prover and verifier do not have to interact over multiple rounds. \textbf{Knowledge} refers to the fact that the statement is true and the prover knows a secret, also called “\textbf{witness}” that establishes that fact. If a SNARK proof allows for proof verification without revealing a witness it becomes a zkSNARK.

Generating a zkSNARK proof is a multi-step process that involves breaking down logic to primitive operations like addition and multiplication, creating an \textbf{Arithmetic Circuit} consisting of gates and wires. This circuit is then converted to a \textbf{R1CS} (Rank-1 Constraint System) which constrains the circuit to verify, for instance, that values are being calculated correctly with gates. The next step converts these constraints in the circuit to a Quadratic Arithmetic Program (\textbf{QAP}). QAP represents these constraints with polynomials rather than numbers. Afterward, together with Elliptic Curve Cryptographic (ECC) and QAP, a zkSNARK proof is formed.

\subsubsection{Trusted Setup Ceremony}

A trusted setup ceremony is a cryptographic procedure that involves contributions of one or more secret values, $s_i$ from one or more parties into the trusted setup system to eventually generate some piece of data that can be used to run cryptographic protocols. Once this data is generated, the secrets, $s_i$ are considered \textbf{toxic waste} and must be destroyed because possession of these makes it possible to rig the system. In the context of ZK, for example, it can be used to generate proofs that are false positives.

There are multiple kinds of trusted setup procedures. The original ZCash ceremony in 2016 \cite{zcash-ceremony} was one earliest trusted setups being used in a major protocol. One trusted setup of particular interest is \textbf{powers-of-tau} setup which is multi-participant and is commonly used in protocols. To get an idea of how multi-step setups work consider the image below where, secrets, $s_1, s_2, ..., s_n$

\begin{figure}
    \centering
    \includegraphics[scale=0.3]{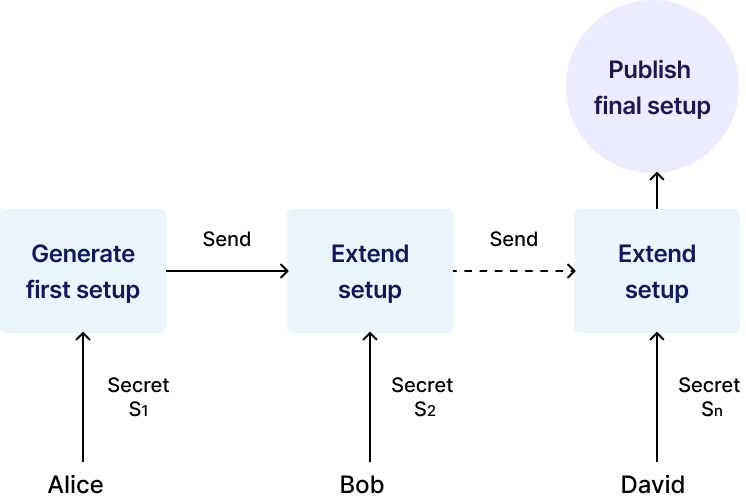}
    \caption{Trusted Setup Ceremony}
    \label{fig:ceremony}
\end{figure}

To be a bit precise, the secrets are used in Elliptic Curve operations for security. An input $s_i$ is used to generate successive points on the curve given the generator $G$ by operation $G*s_i$. The result of the contribution, $G*s_i$ is fed to the next step. The next participant cannot determine the previous contributor's secret because it is a discrete log.

A more detailed discussion can be found at \cite{trusted-setup}.

\subsubsection{Groth16}
\textbf{Groth16} is a SNARK proving system proposed by Jens Groth \cite{groth16} in 2016. It is a non-interactive proof system that is commonly used in privacy protocols and is fast with small proofs. The prover complexity is $O(n_c)$ where $n_c$ is the number of rank-1 constraints.

Groth16 requires a trusted setup to generate proving keys and verifying keys. The setup is required to make the proofs succinct and non-interactive.

\subsection{Merkle Tree}

Merkle Tree is a data structure that is used to store and verify the integrity of data. It is a binary tree where each node has two children. The leaves of the tree are the data blocks that are stored and the internal nodes of the tree are the hashes of children below them. Consequently, the root of the tree is the culmination of all data blocks in the tree. This means that if any data blocks at the leaf nodes are changed or corrupted it will propagate to the root which will change too.

\begin{figure}[h]
    \centering
    \includegraphics[scale=0.15]{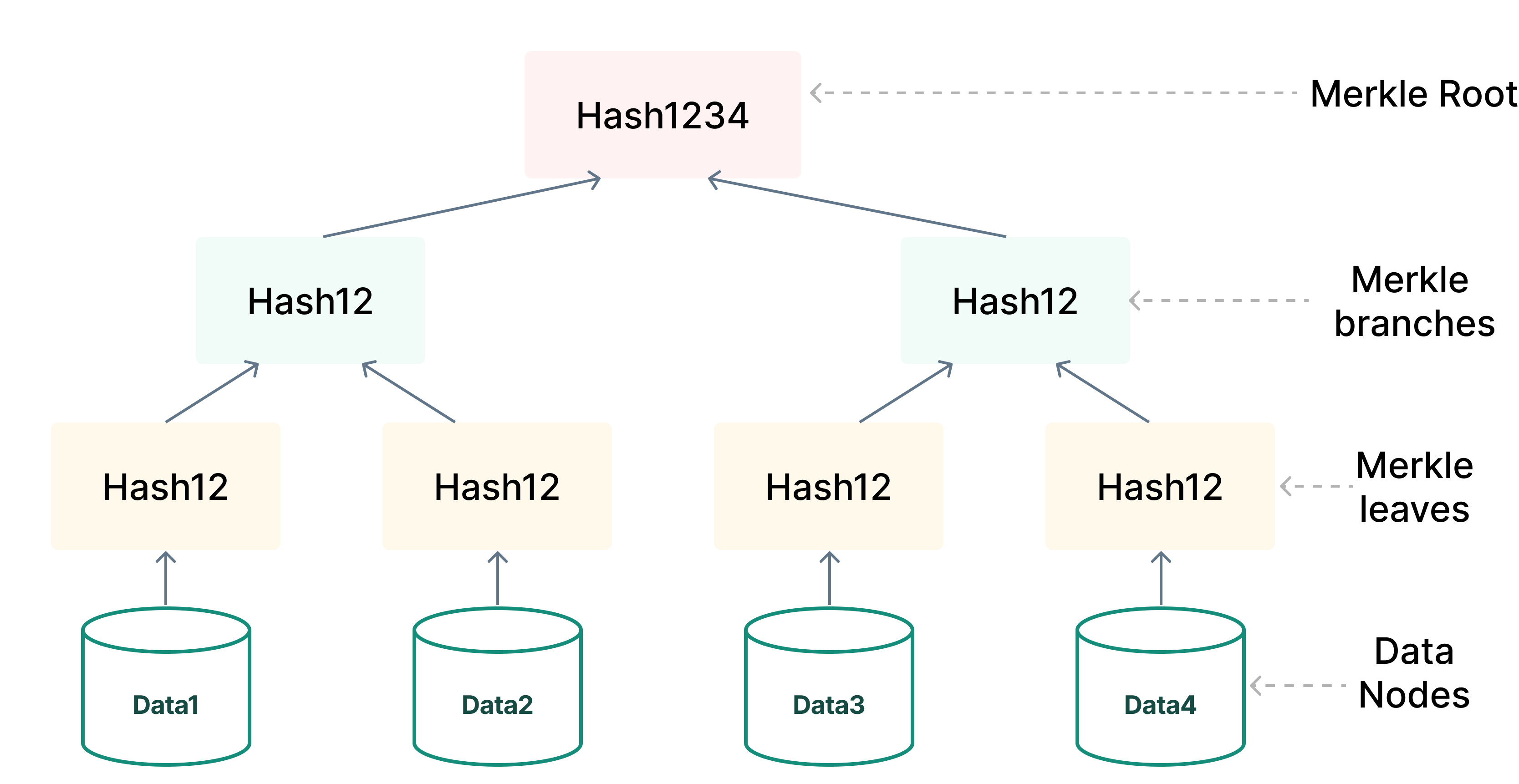}
    \caption{Merkle tree}
    \label{fig:merkle}
\end{figure}

The root hash of the Merkle tree can be compared to a reference to verify the integrity of the data. This makes the Merkle trees very efficient.

A merkle tree is maintained in zkFi to record user assets. User-deposited assets are represented in the form of a hash called \textbf{commitment} that gets stored in a Merkle tree structure. During withdrawal, the same user submits a proof a.k.a \textbf{Merkle Proof} (among others) of inclusion of the same commitment hash in the tree and proof of knowledge of the hash's pre-image, without revealing any traceable information like an index of commitment in tree or constituents of pre-image.

\section{Conclusion}
For web3 to reach mass adoption, privacy is an unquestionable facet. While zero-knowledge technology has solved the privacy problem, earlier solutions have clearly indicated that strong compliance to prevent illicit use cannot be ignored at all. Furthermore, to encourage protocols and developers to build privacy into their products, infrastructure solutions must be present that are easy to integrate and build on top of.

In this paper, we demonstrated that an infrastructure-based privacy solution with built-in compliance is possible. We proposed a middleware solution through an SDK package that is easy to integrate and abstracts away solutions to privacy and compliance without sacrificing developer experience. Consequently, rendering the protocols and developers the freedom to innovate and focus on solving their core problem instead of worrying about user privacy or compliance.

\end{document}